\documentclass{PoS}

\newcommand{\lsp}{LS~I~+61$^{\circ}$303}
\newcommand{\lsi}{LS~I~+61$^{\circ}$303~}

\newcommand{\vt}{V773\,Tau\,A}

\newcommand{\grs}{GRS 1915+105~}
\newcommand{\beq}{\begin{equation}}
\newcommand{\eneq}{\end{equation}}

\title{Transient sources at the highest angular resolution}

\ShortTitle{Transient sources at the highest angular resolution}

\author{\speaker{Maria Massi}\\

        Max-Planck-Institut f\"ur Radioastronomie, Bonn, Germany\\
        E-mail: \email{mmassi@mpifr-bonn.mpg.de}}

\abstract{
By definition transients are  sudden events, some,  like supernovae,
are  catastrophic, while  others  might be 
due to recurrent  phenomena.
The aim of  studying   transients  is  to reveal   
the physical conditions causing them,
in this sense ideal targets for monitoring are transients in binary systems.
In these systems  the physical process responsible for the transient
depends directly or indirectly
on the interaction of the two components of the system.
Here I report  on transients in stellar binary systems
at  two extremes
    of stellar evolution:    
    a T~Tauri system formed by two young  low mass stellar objects,  and   
    X-ray binary systems formed by a star and  a neutron star or a black hole,
   i.e., end points in the life of  massive stars. 
    VLBI observations of the young binary system \vt{} resolve the binary separation 
and can be overlapped with the optical frame. 
Consecutive VLBI observations  showing the evolution of the radio emission
with respect to the two stellar objects are an unvaluable tool  
for a better understanding of the magnetic field topology of T Tauri stars. 
The characteristics of radio jets in X-ray binaries  are summarised and compared with
those of the radio emission of the gamma-ray binary \lsp. 
Timing analysis of radio and \textit{Fermi}-LAT observations provide  constraints for theoretical models
that can be tested  by VLBI observations.
}

\FullConference{12th European VLBI Network Symposium and Users Meeting,\\
		7-10 October 2014\\
		Cagliari, Italy}

\begin{document}

\section{Introduction}
Radio  transients [1] are observed  from
the  nearby  Sun to  cosmological distances.
A classification of transients following the physical process in act is  complicated:
different physical processes, as    magnetic reconnection, shocks, electron beams, etc.,  may work in the same object.
A possible classification is that  based on the kind of  astronomical objects where the transient occur:
    cool stars,  supernovae,  pulsars and  jets.
Straightforward and based in  differences in timescale 
is the classification of  radio transients in fast transients and slow transients.
Fast radio transients may have timescales of nanoseconds  to minutes 
and they typically are discovered in time-series data.
The slow radio transients,  
variable on timescales of seconds up to years,
are typically discovered in images and are those described here.
Transients in binary systems, where 
  the interaction between  the  components is  the cause of  the transient,  
  are  ideal targets for monitoring studies.
In the next section I will therefore examine slow radio  transients in binary systems.
\section{Transients in a young-stellar system}
The  binary system V773 Tau A [2] shows outstanding
magnetic activity 
 demonstrated by
radio and X-ray flares and the presence of
large, cool, photospheric spots.
A  periodicity  of 51.1 d, the same  as the orbital one,
has been discovered 
 in the radio flaring activity
via long-term monitoring with
the Effelsberg 100-m  telescope:
 large flares cluster around periastron passage [3].
The orbit, with  an eccentricity of $e\,=\,0.27$ [2], results
at apastron in  a distance of 52 stellar radii (1 $R_*$ = 2 solar radii)
between the two stars,
and  at periastron of 30 stellar radii.
If the  strong  flaring activity at periastron  is due to interacting coronae,
then the magnetic structures should have sizes of at least 15 stellar radii.
Observations  at 90 GHz around  periastron have monitored the onset and decay of a large flare
and show that the decay of few hours is consistent with continuous  leakage of the emitting relativistic
electrons
from a  magnetic structure of 10--20 stellar radii [4].
There exists  a close relationship between flares and interaction of magnetic structures.
In fact, as  observed on the Sun, flares can be  triggered by
interactions between new and older emergences of magnetic flux
in the same  area [5].
Part
of the
magnetic energy released during reconnection goes to accelerate
a fraction of the thermal electrons  trapped in the flaring coronal loop
and a  population of relativistic electrons  is
produced [6,7].
\begin{figure}[h]
\center
\includegraphics[width=.15\textwidth]{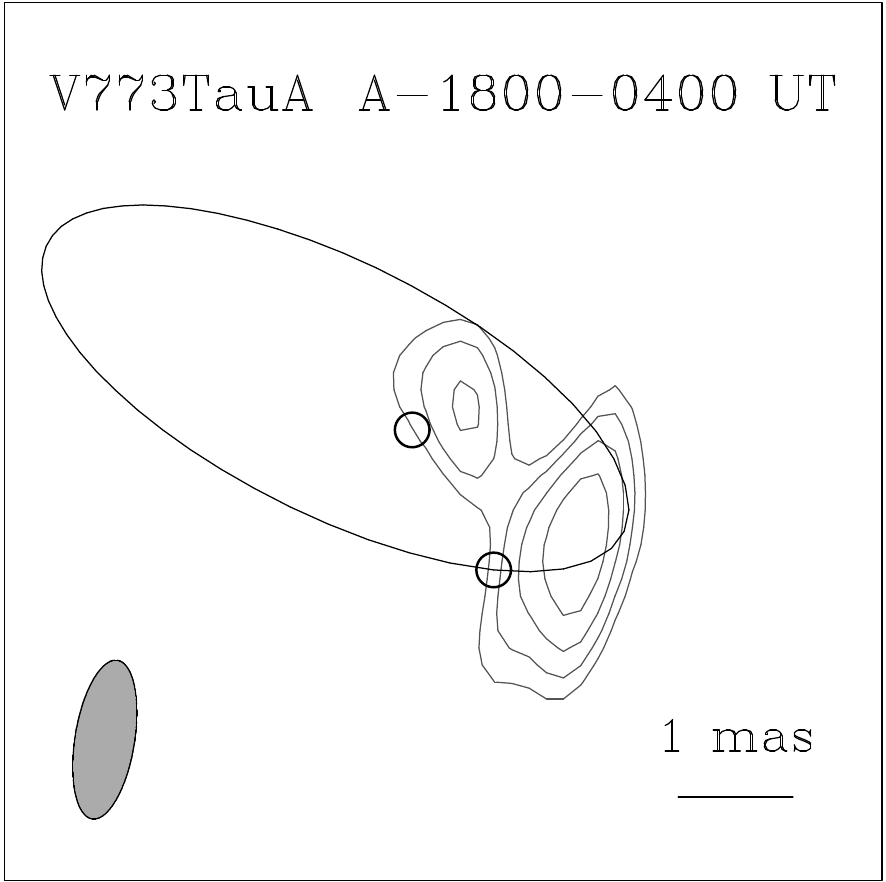}
\includegraphics[width=.15\textwidth]{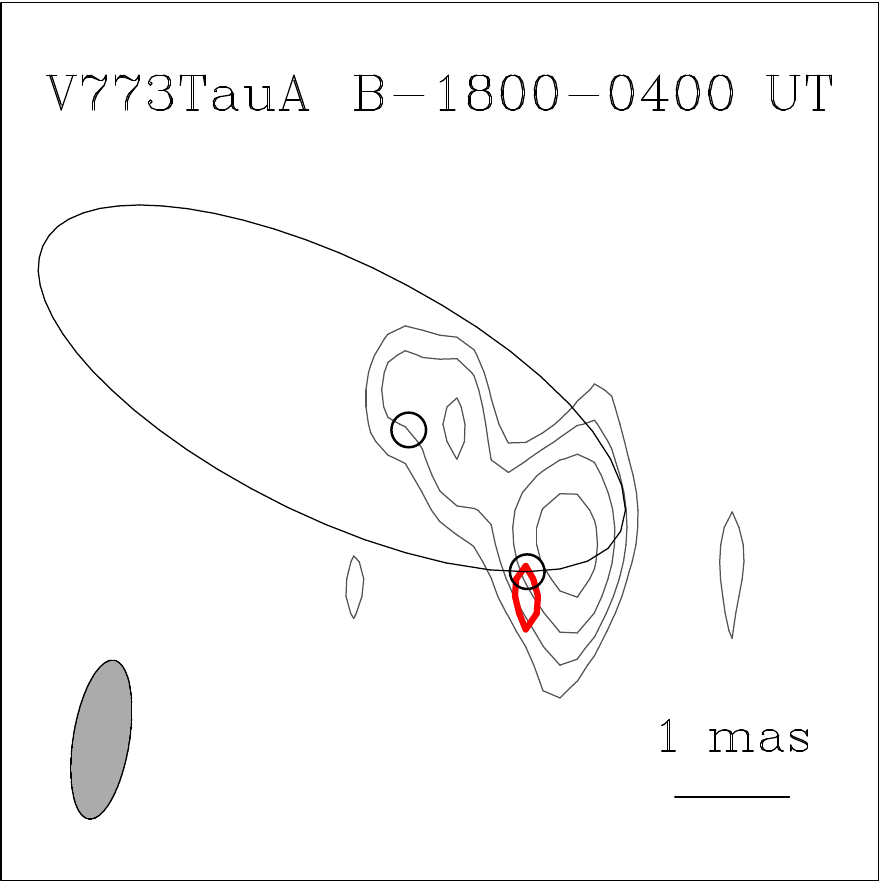}
\includegraphics[width=.15\textwidth]{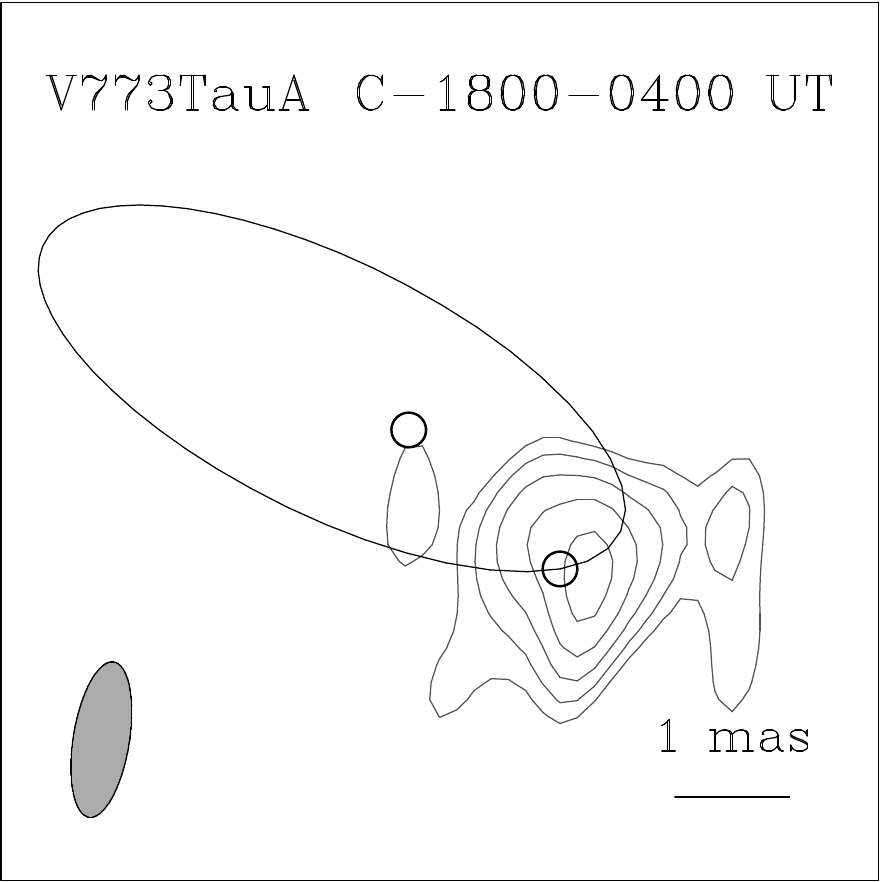}
\includegraphics[width=.15\textwidth]{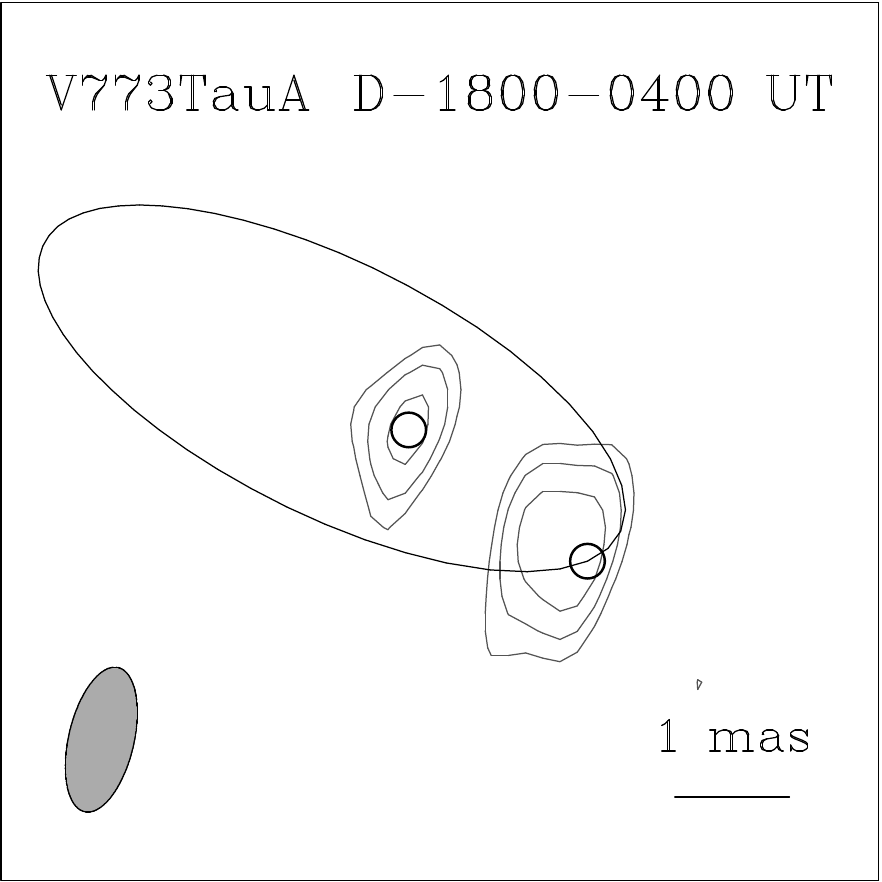}
\includegraphics[width=.15\textwidth]{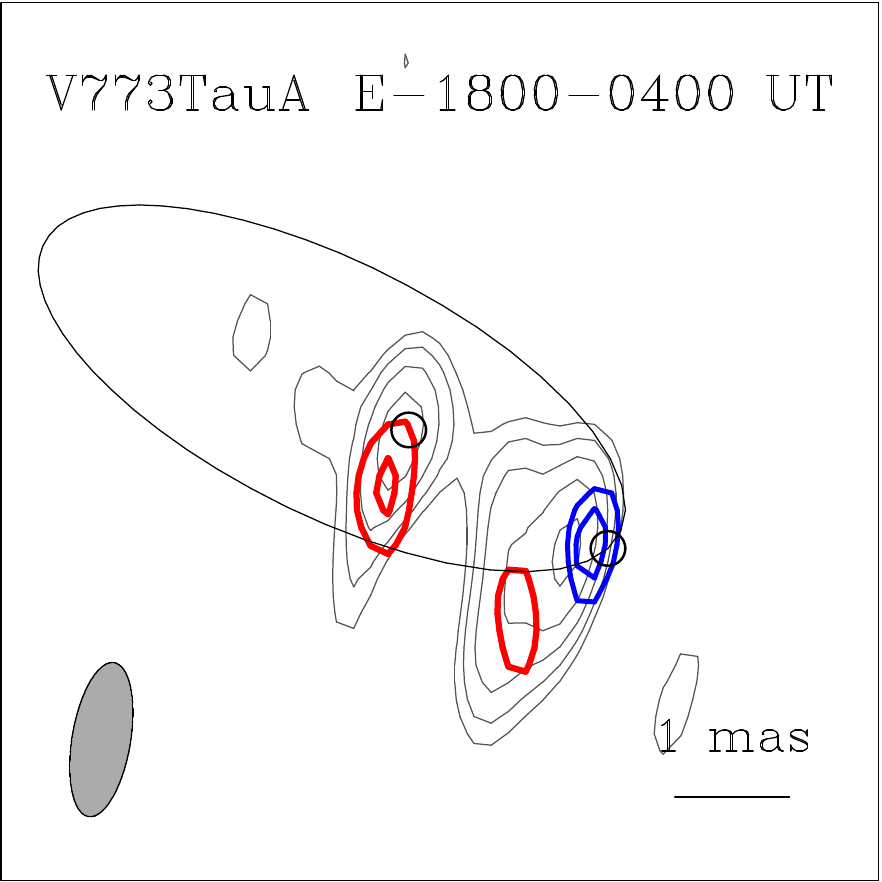}
\includegraphics[width=.15\textwidth]{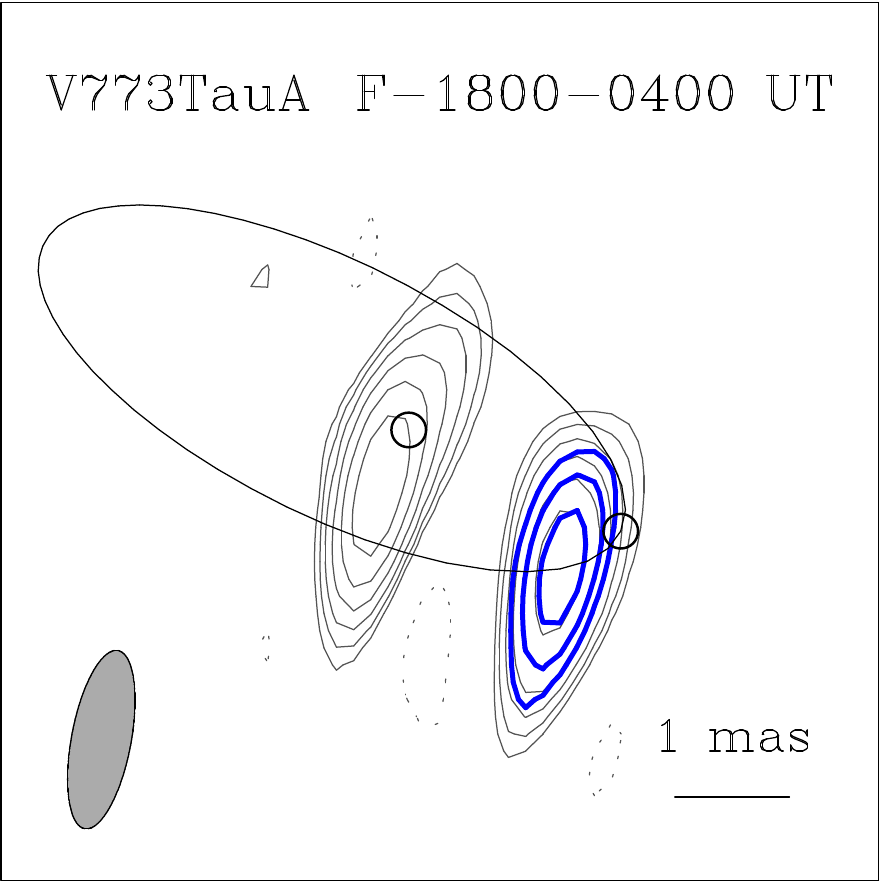}
\caption{
Phase-referenced images of consecutive (25 February $-$  2 March 2010)  8.4 GHz VLBA+EB observations of V773 Tau A.
Overlaid is the orbit of the binary stellar
system; the circle of 4 R$_*$ centered at each star's position indicates a
symmetrical corona of an average size of 3 R$_*$ (the stellar radius corresponds to 0.07 mas at 132.8 pc distance).
The restoring beam is shown in the bottom left corner of each map; it is on average
of (0.5 $\times$ 1.4) mas.
Circularly polarized emission (traced in colour) is associated with the 
secondary star in E and  F with V/I equal to $-4.3\% \pm 0.4\%$, $-8.0\% \pm  0.3\%$  and also to the
primary in E, with opposite polarization sense and V/I of $4.0\% \pm 0.4\%$ 
[8].}
\label{fig1}
\end{figure}
These   relativistic particles
gyrating around the magnetic field lines of the  loop,
where they are confined, generate  synchrotron emission in the radio band.
Applying this knowledge of solar flares to V773 Tau A,
it is clear that in this system the observed  relationship  between  intensity of the flare
occurrence and the distance between  the two stars
indicates another, new mechanism of magnetic interaction, that of interacting coronae.
In this case  magnetic reconnection
would take place far out from the stellar surfaces,  where the two coronae interact
with each other.
Figure  1 shows phase referenced maps of consecutive  VLBA+EB observations aimed to
spatially trace the flare evolution and  polarized emission
around  periastron passage  where the intensity of the flares is highest [8].
The visual orbit  is derived from an 
interferometric-spectroscopic orbital model [2]
and  the primary star is overlapped with the North-East feature in the map of run D.
By using the images, the total flux density monitoring,  model fitting and the information from circular polarization (here in colour) 
we have a powerful tool for a better understanding  of 
the physical process triggering the transient 
and  of the 
magnetic field topology of  weak-lined T Tauri stars.
\section{Radio transients in X-ray binaries} 
X-ray binaries are stellar systems formed by a compact object (black hole or neutron star)
and a normal star.
Neutron stars in X-ray binaries may have quite different values of 
 magnetic field, being the range of B about $10^{8}-10^{12}$ G;
however,  X-ray binaries with a radio emitting jet, systems called "microquasars",
have as  compact object  either an accreting black hole or an accreting  neutron star 
with low magnetic field ($10^{8}$ G) [9].
Radio jets have been imaged at high resolution for the three neutron star systems: Scorpius X-1 [10],
Circinus X-1 [11] and Cygnus X-2 [12]. 
The statistically more powerful  radio 
jets associated to accreting black holes
have quite  well studied spectral characteristics.  
There are two kind of jets; the steady jet and the transient jet [13].
The origin of a steady jet is relatively well understood as a result  of magneto-rotational  
processes: an initial vertical magnetic field treading the accretion 
disk is bent by the  differentially rotating disk, then
  magnetic pressure gradient  accelerates  plasma out of system and 
  magnetic tension   pinches and collimates the outflow into a jet [14].
Changes in the plasma  density  
and in the strength of the magnetic field along the conical jet create distinct regions of
 synchrotron emission with each region contributing  with a  spectrum peaking at a
 different frequency.
 Then the overall
 spectrum, observed with a spatial resolution insufficient to resolve
 the individual parts of the jet, will be almost flat,
 i.e., with a  spectral index $\alpha \sim 0$,  or inverted, i.e.,  $\alpha > 0$ [13, 15]   
(flux density $S\propto \nu^{\alpha}$). 
The  other type of jet associated to microquasars is
the so called transient jet.
\begin{figure}[h]
\center
\includegraphics[width=.25\textwidth]{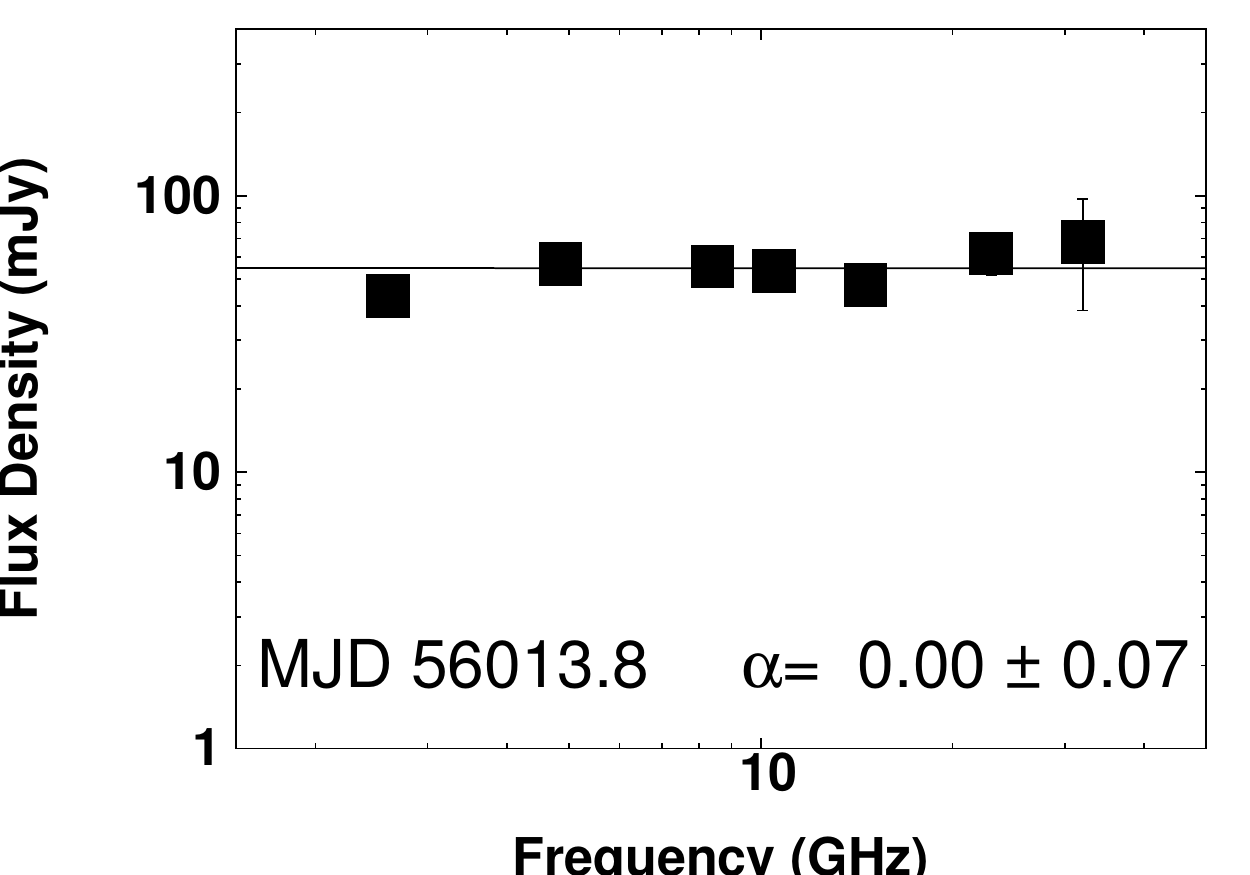}
\includegraphics[width=.3\textwidth]{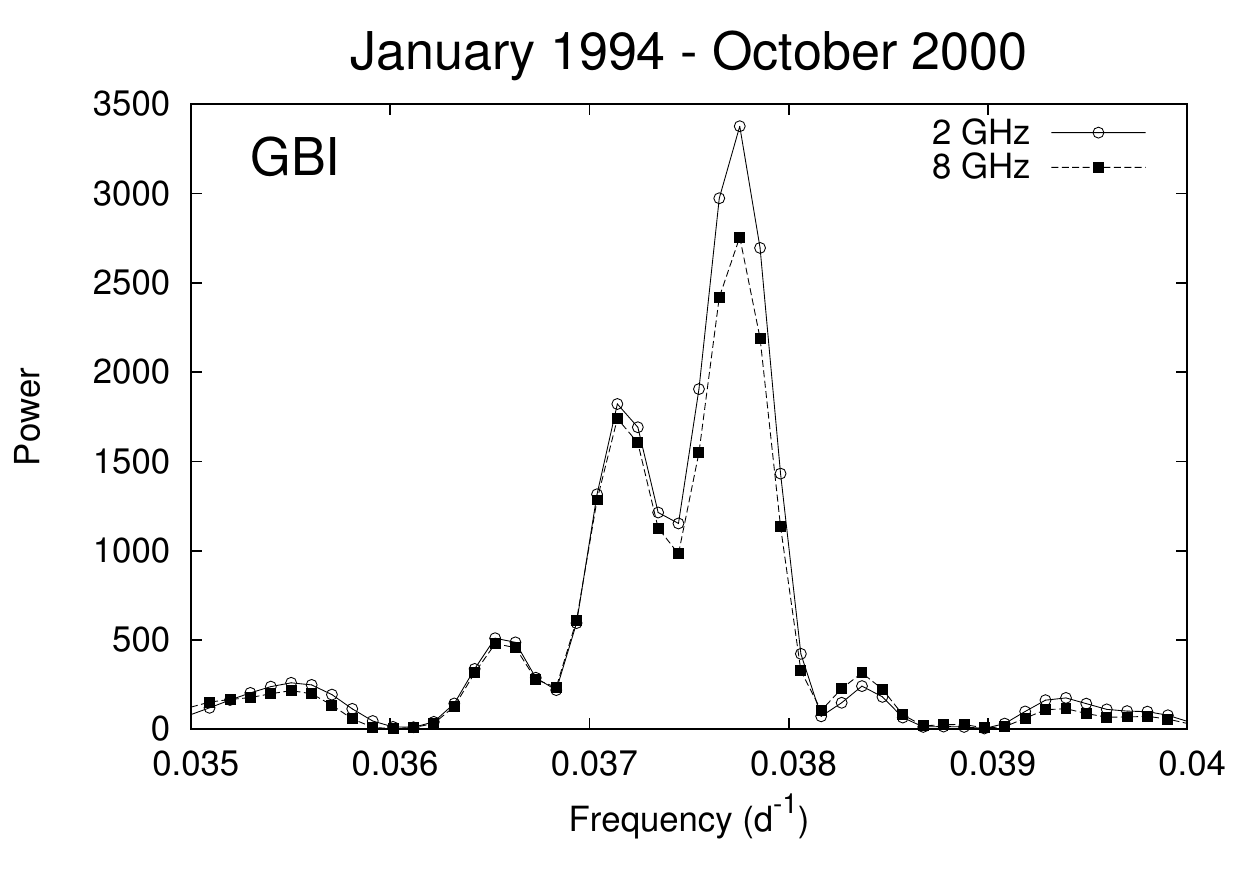}
\includegraphics[width=.3\textwidth]{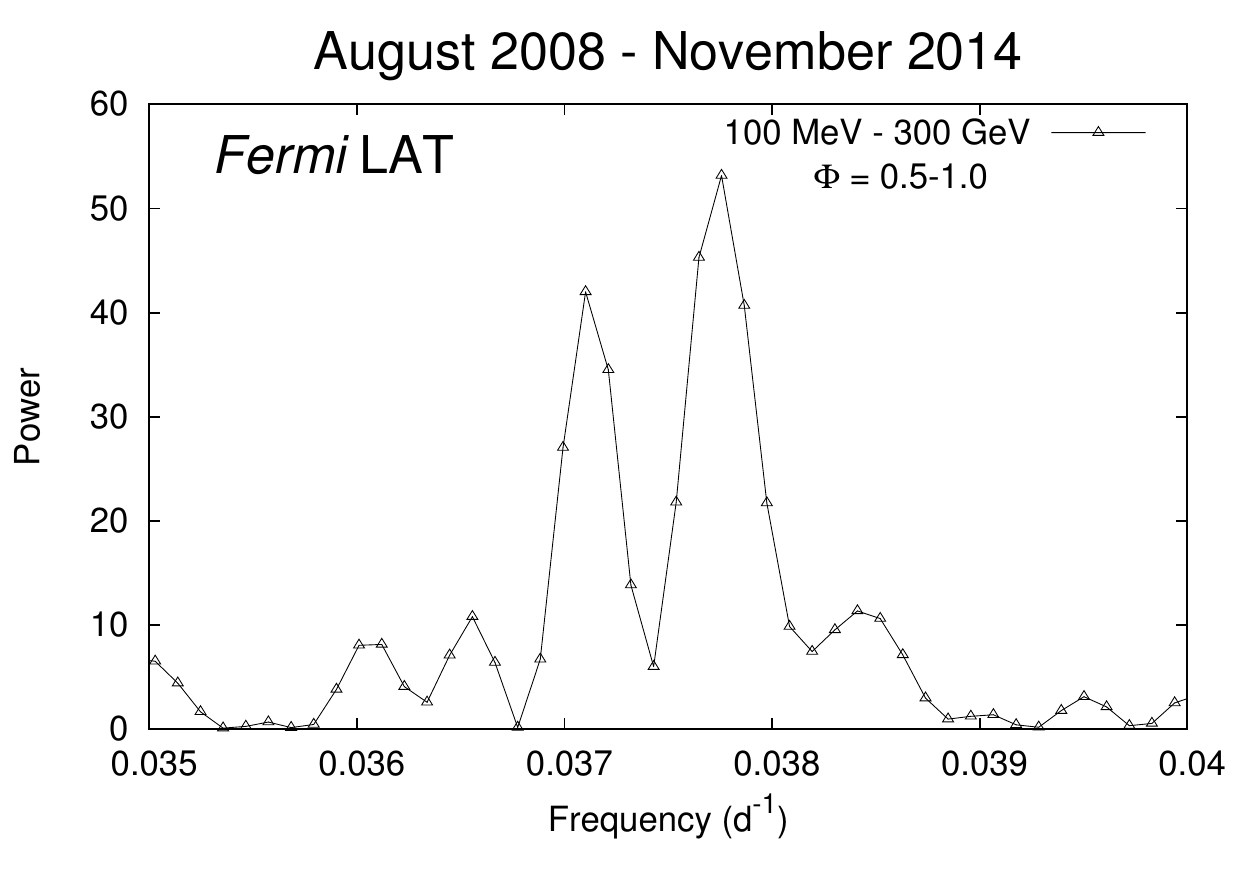}
\caption{Left: Flat radio spectrum of \lsi measured with the Effelsberg 100-m telescope
at 2.8, 4.85, 8.35, 10.45,  14.3, 23 and 32 GHz [24]. Center: Periodogram of GBI radio data. Right: Periodogram of 
\textit{Fermi}-LAT data  (see Sect. 3.1).}
\label{fig2}
\end{figure}
It corresponds to an optically 
thin  radio  outburst (i.e.,
$\alpha < 0$) and  occurs  always {\it after} the flat spectrum phase.
The transient jet is thought to be associated 
to  shocks where  high relativistic plasma 
catches up with the pre-existing slower-moving material of the steady jet  [13].
The switch from the self-absorbed jet to  the transient one presents differences 
in the different sources.
In the microquasar \grs the switch is between a steady,  plateau  state
and the optically thin outburst [Fig. 1,  in 16], whereas in XTE J1752-223 the switch is between
an optically thick outburst  
 and an  optically thin outburst [Fig. 1, in 17].
A unique case of a radio outburst from an X-ray binary having as compact object a
non-accreting neutron star  is PSR B1259$-$63.
The neutron star is in this case a young pulsar, i.e., having both 
a  strong B ($10^{12}$ G) and a fast rotation (msec period). 
Around periastron passage the   interaction of the  pulsar wind  
and  the  wind of the companion star generates  an optically thin outburst [18]  
mapped at high resolution [19].
\subsection{Latest results on a transient source: The gamma-ray binary \lsp}
The stellar system \lsi is formed by a compact object and a Be star
in an eccentric orbit, $e = 0.72 \pm 0.15$ [20].
High resolution radio images
 show  a structure 
that  not only changes
position angle, but it is even sometimes one-sided and at other times
two-sided.
 This suggested the hypothesis of \lsi being a precessing
 microquasar [21].
 A precession of the jet leads
to a variation in the angle between the jet and the line of sight, and
therefore to variable Doppler boosting. The result is both a
continuous variation in the position angle of the radio-emitting
structure and its  flux density.
The rapid variations were alternatively interpreted [22]
to be due to a young pulsar whose wind enters in collision with the disc/wind of the companion star,
i.e., a system as PSR B1259$-$63.
\begin{figure}[]
\center
\includegraphics[width=0.23\textwidth,angle=-90]{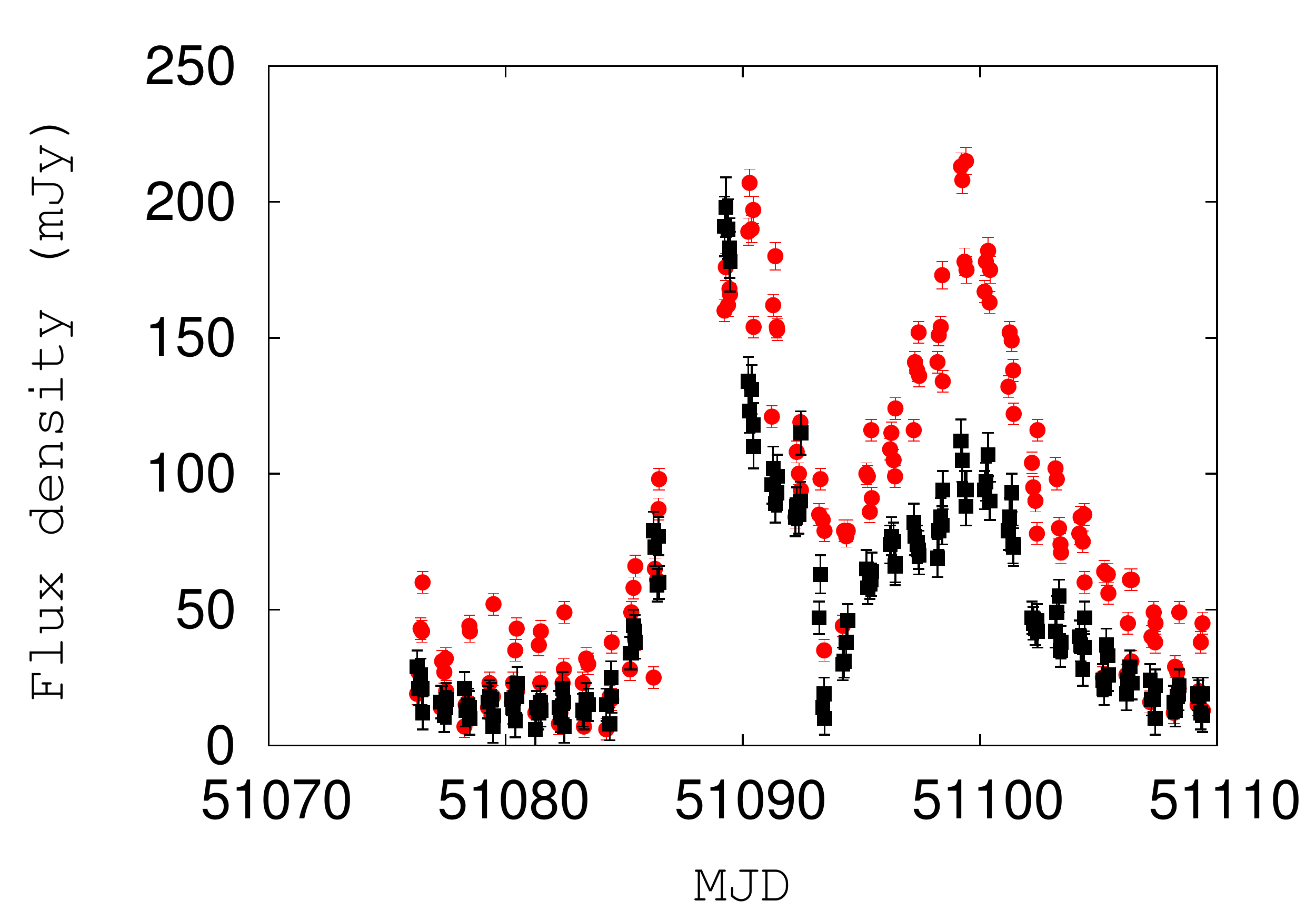}
\includegraphics[width=0.23\textwidth,angle=-90]{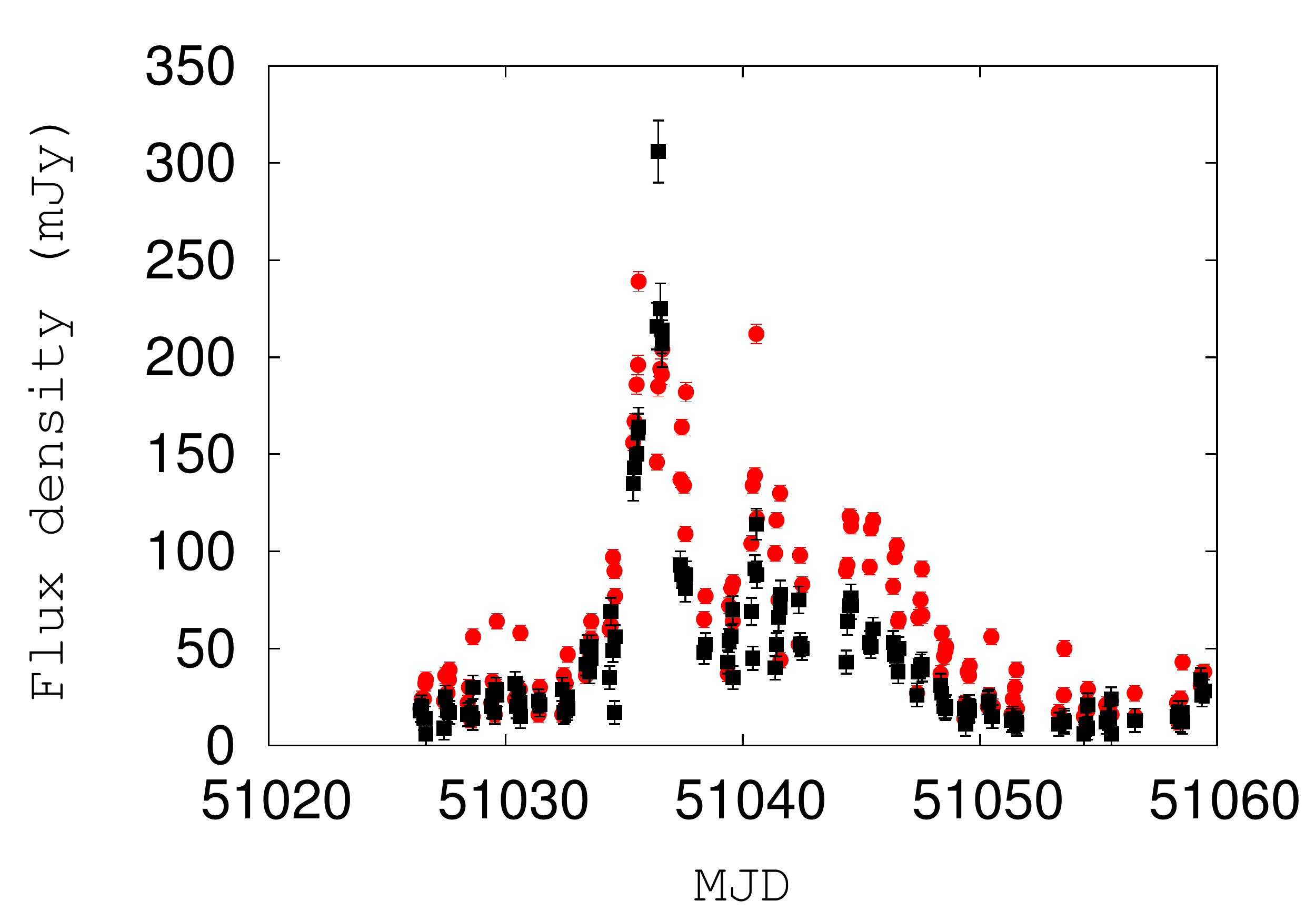}
\includegraphics[width=0.23\textwidth,angle=-90]{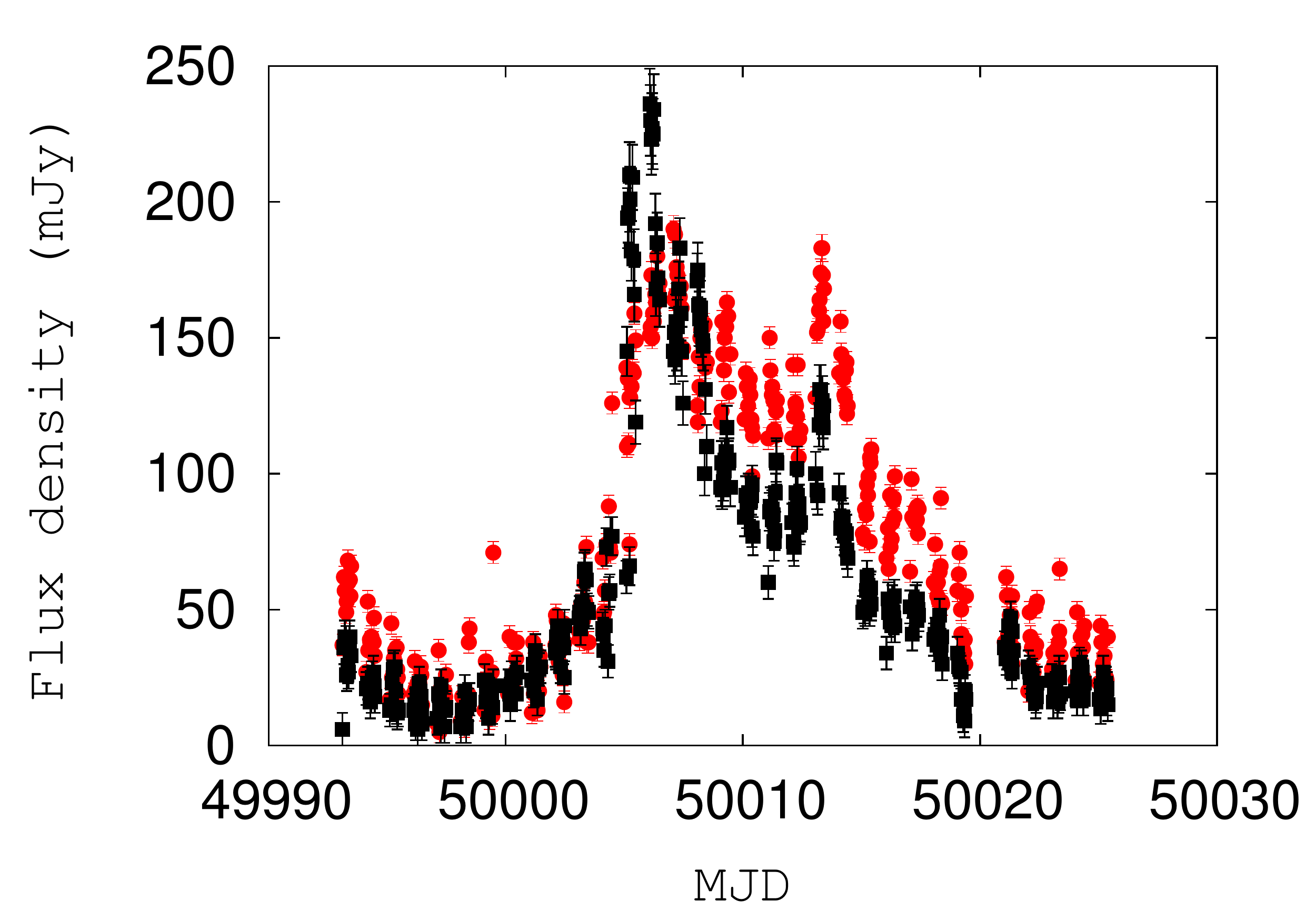}\\
\includegraphics[width=0.23\textwidth,angle=-90]{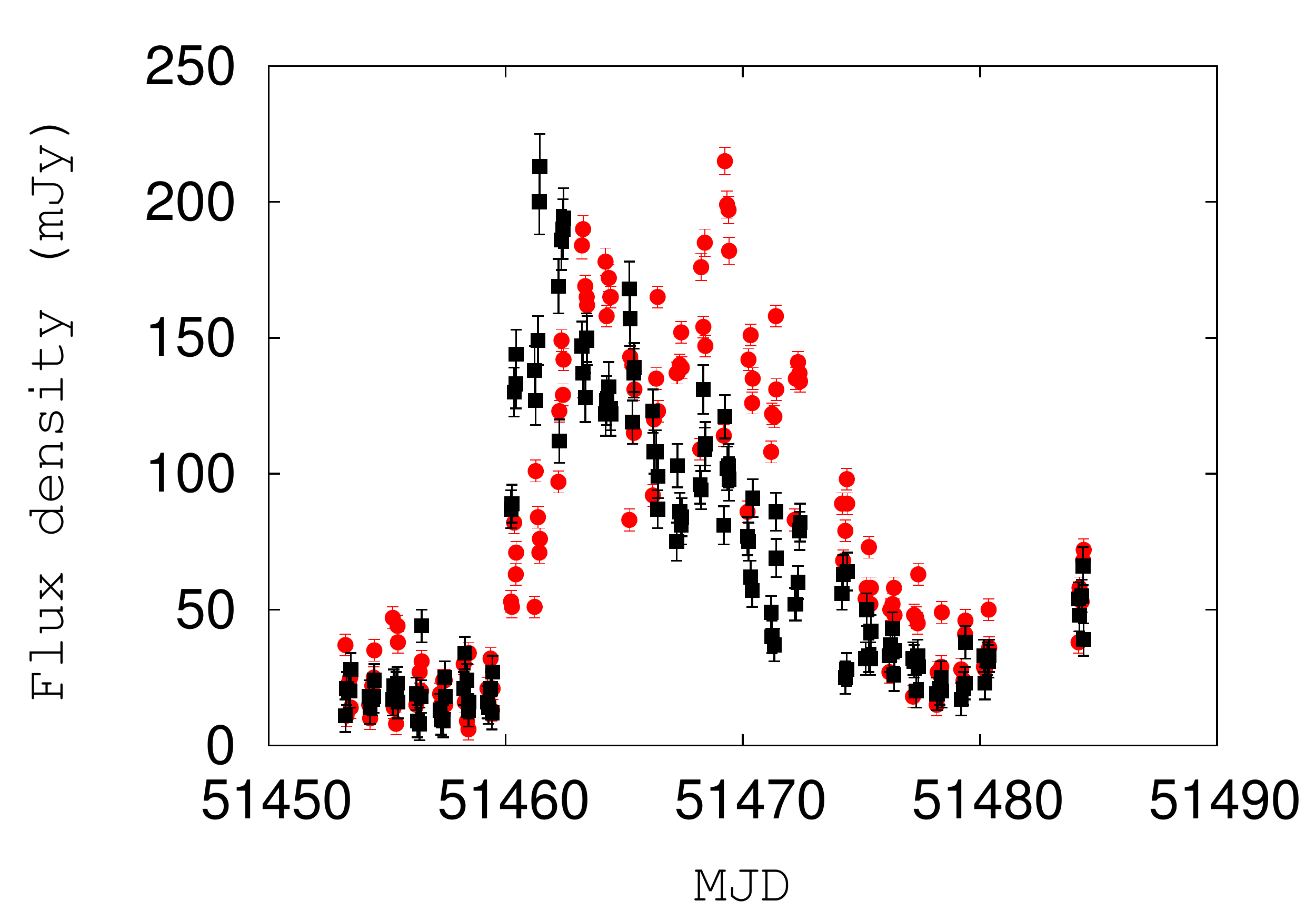}
\includegraphics[width=0.23\textwidth,angle=-90]{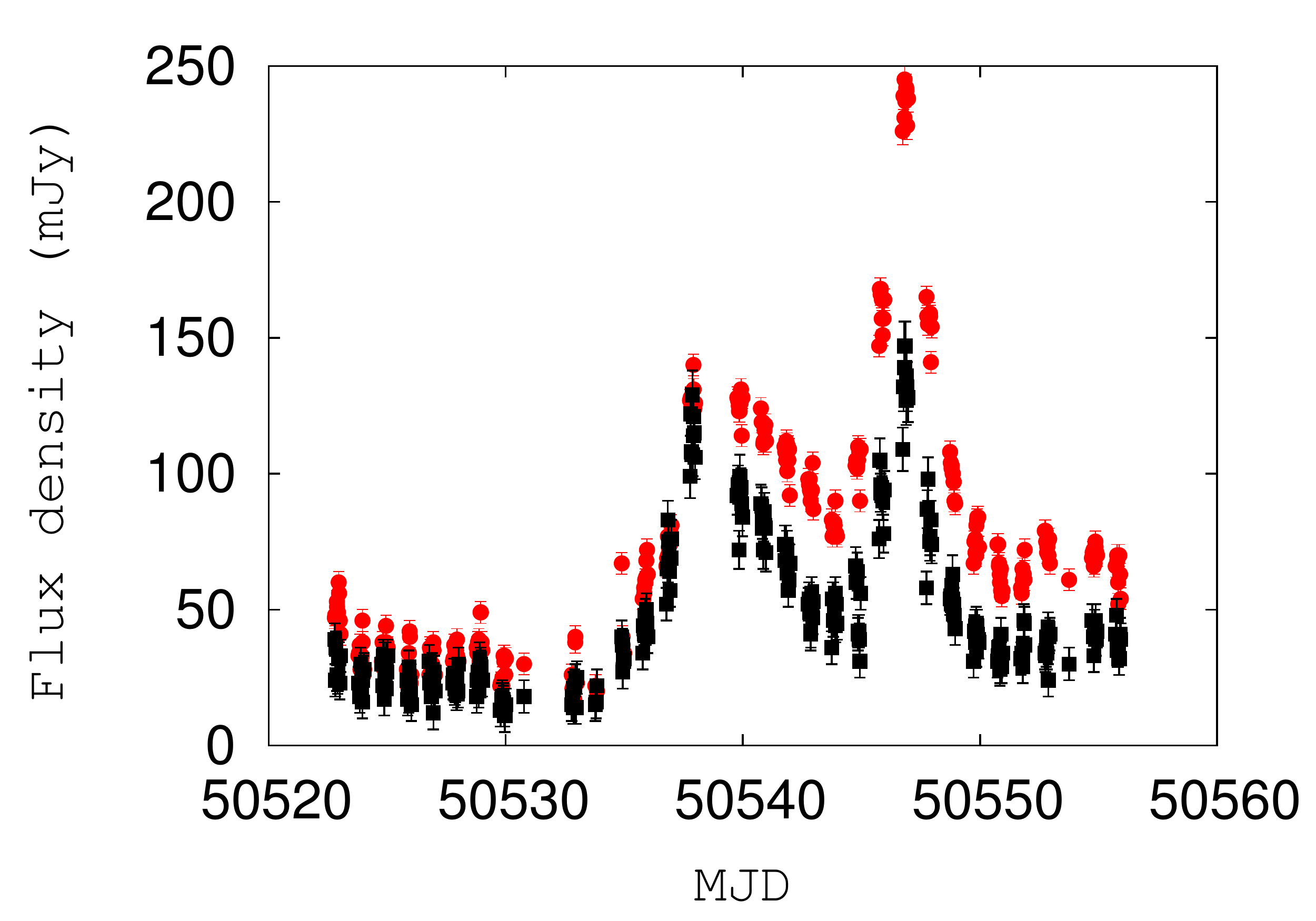}
\includegraphics[width=0.23\textwidth,angle=-90]{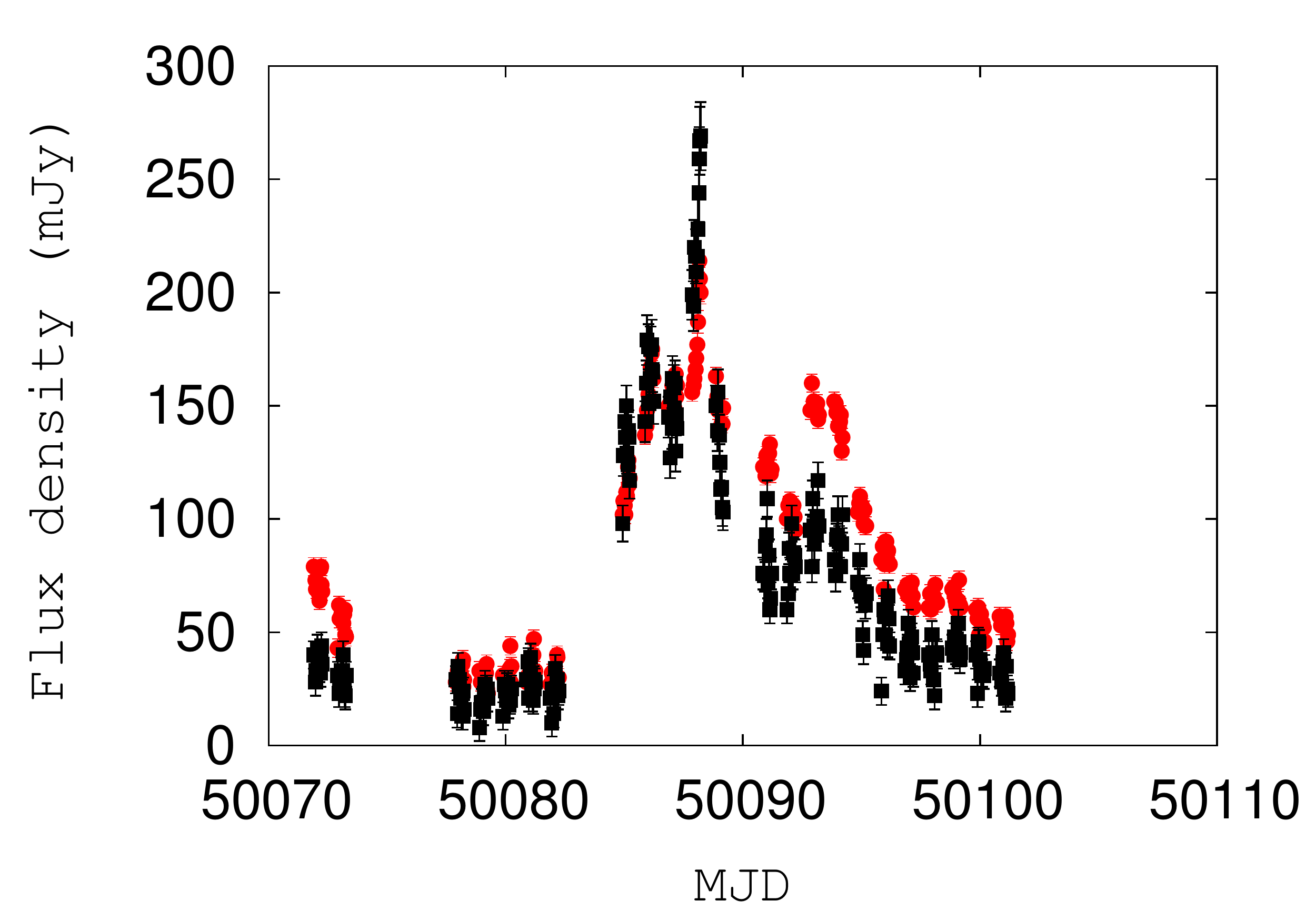}\\
\caption{Light radio curves  of \lsi at 2 GHz (red) and 8 GHz (black) 
	GBI data vs time (MJD) [25]. An optically thin outburst follows an  optically thick outburst as in 
 XTE J1752-223 (see Sect. 3).}
\end{figure}
As we saw in the previous section, the radio spectral index for a radio emitting jet of a microquasar
and that of a radio emitting nebula associated to interacting winds are quite different.
A flat/inverted spectrum is expected for a radio jet, followed by an optically thin outburst
in case of a transient. Just an optically thin outburst is expected for the pulsar nebula.
It is therefore worth to examine the radio characteristics of \lsi in detail.
\begin{figure}[t]
\center
\includegraphics[width=0.35\textwidth]{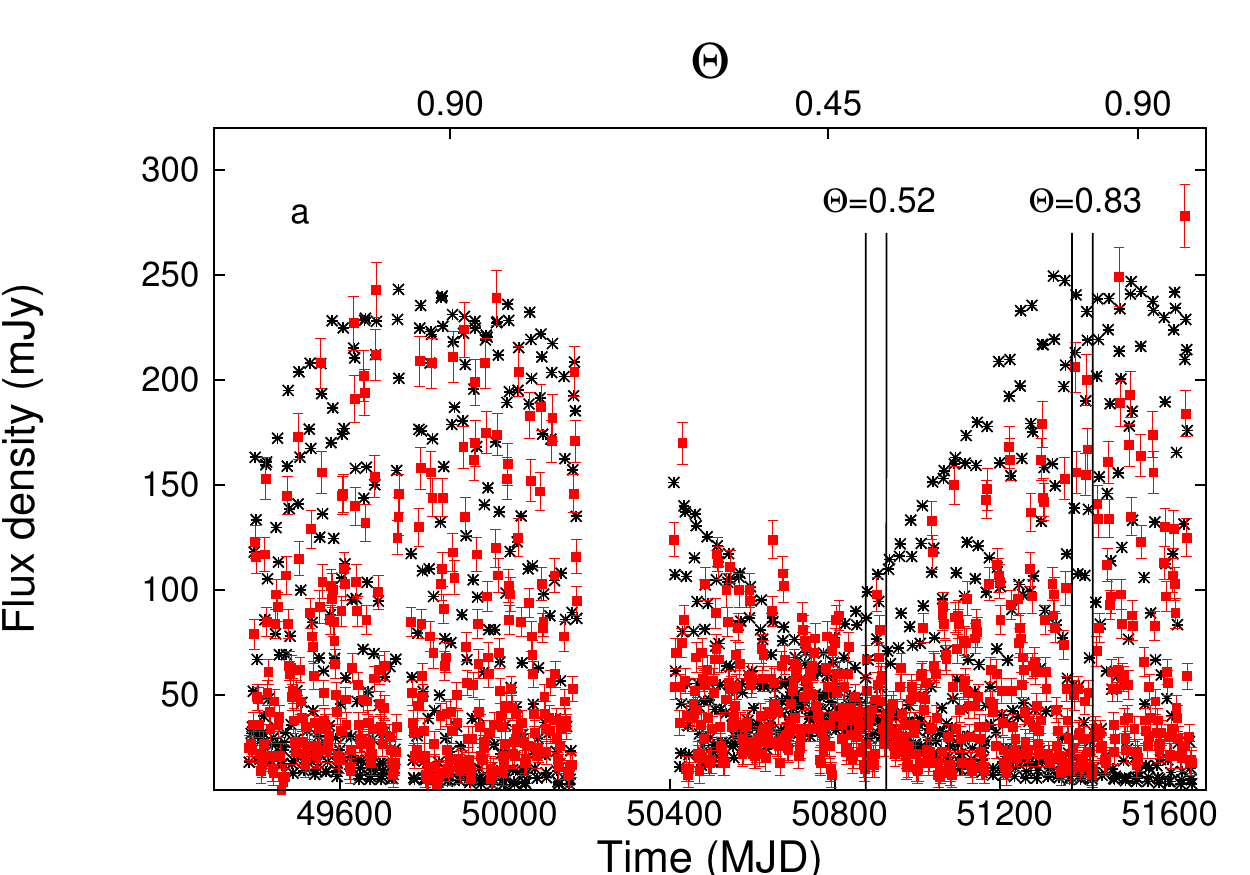}
\includegraphics[width=0.25\textwidth]{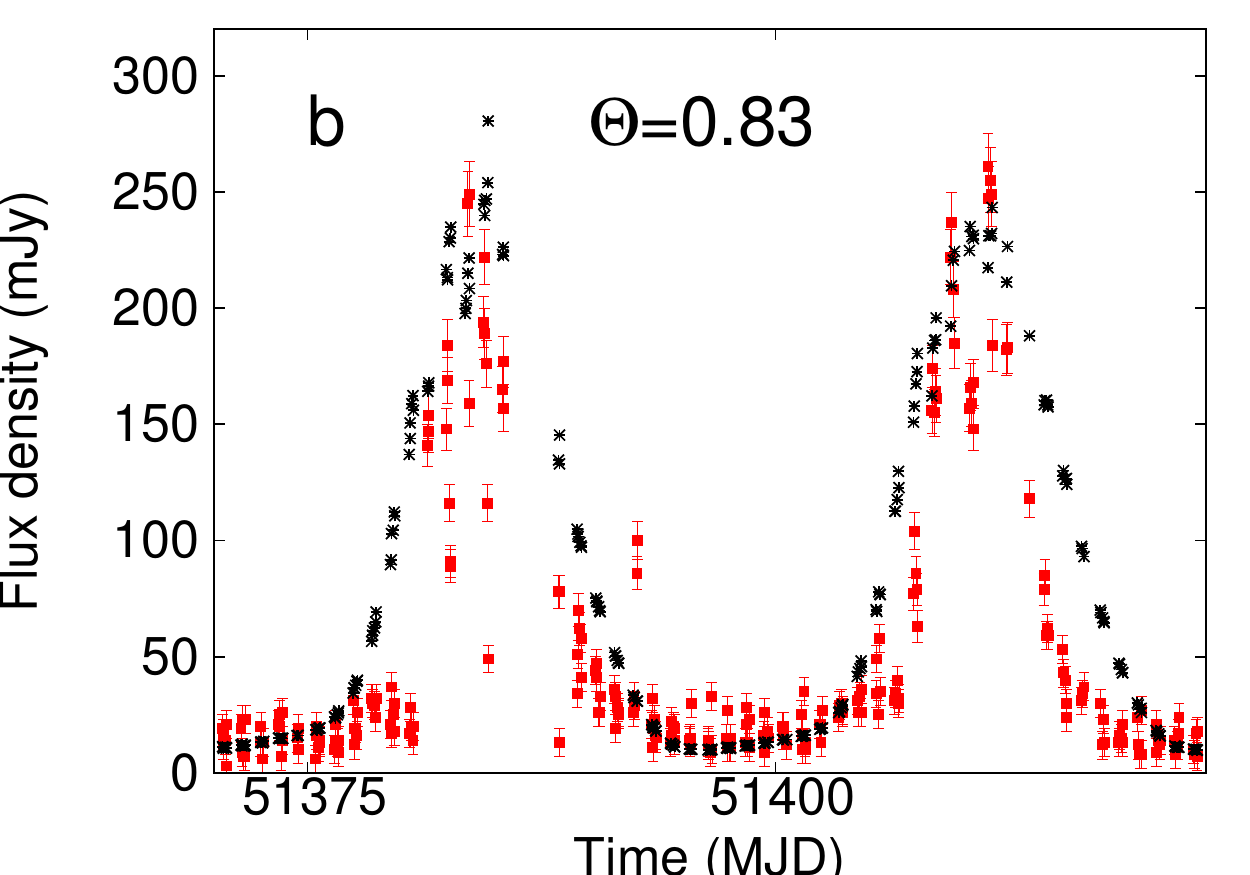}
\includegraphics[width=0.25\textwidth]{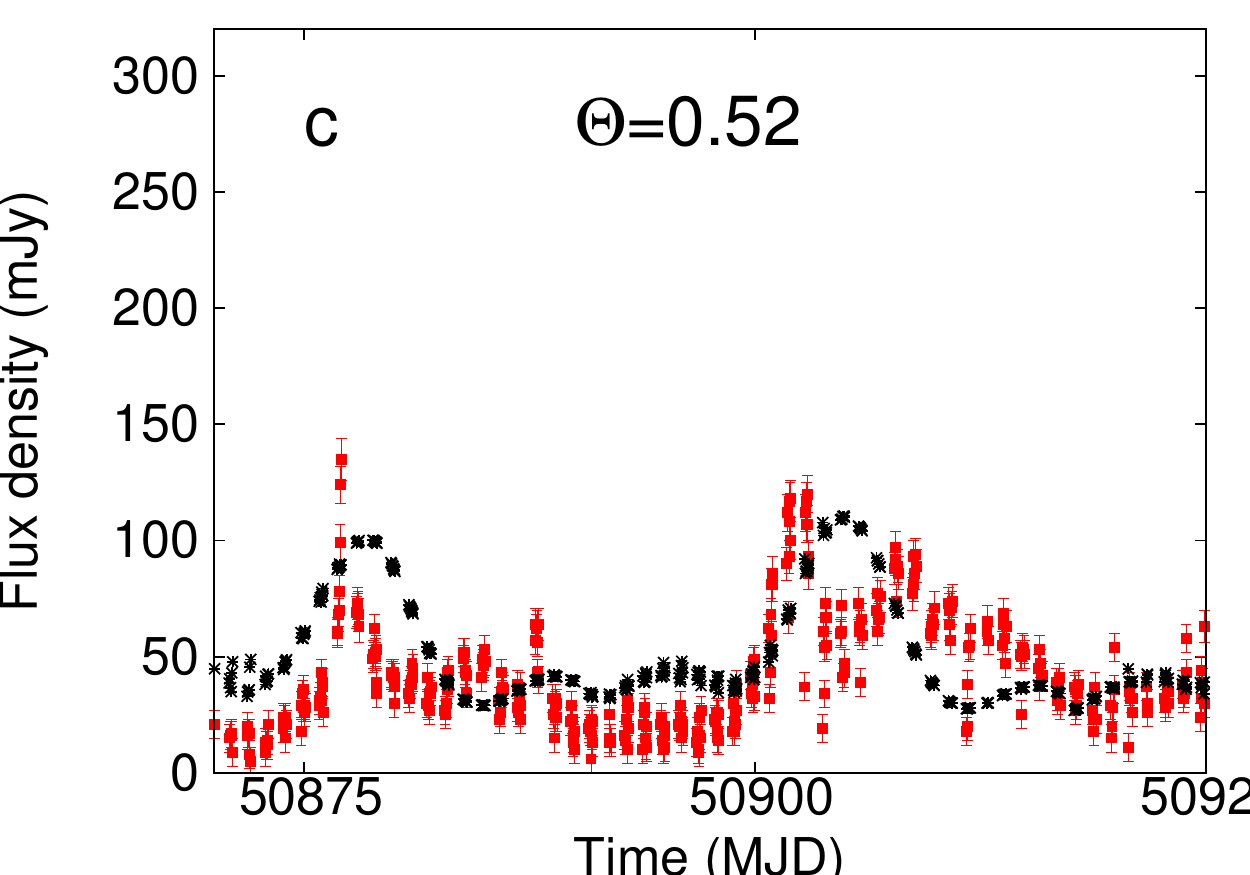}
\caption{Long-term modulation in LSI+61303. 
	a) 8 GHz GBI observations (red) and model (black) data [32]. 
	b and c) Zoom of two intervals of Fig. 4 a.
}
\end{figure}

VLA observations [23] measured a flat spectrum extending from 1.5 GHz to 22 GHz. 
Effelsberg 100-m telescope observations (Fig. 2), reveal  a flat spectrum up to 32 GHz [24].
Observations at two frequencies, 2 GHz and 8 GHz of the Green Bank Interferometer (GBI) 
are given  in Fig.~3. 
As in Fig. 4 of   [25],
it  is  also evident in Fig. 3 that in \lsi    two consecutive outbursts occur: 
there is a first outburst either dominating
at 8 GHz (i.e., with inverted spectrum, $\alpha > 0$) or with flat spectrum
(comparable flux at the two frequencies) 
and then there follows  another outburst, this one clearly dominating at 2 Ghz (i.e., $\alpha <0$).
The system \lsi does not only show the 
self-absorbed jet 
but in addition  there it follows a  transient jet, as  typical for microquasars. 
Finally, following the hypothesis that  a short X-ray burst observed in the direction of \lsi could be attributed
to this system and not to the other X-ray source that is located in the  same field of view,
possible  implications for having a magnetar in a binary system have been analysed [26].
The presence of another X-ray candidate on the  one hand and  the radio observations 
well consistent with those of microquasars  on the other hand do 
corroborate  the hypothesis of   \lsi to  be a  precessing microquasar. 

Deriving the precessional period from  radio images
is not straightforward, because  the radio structures   reflect  the variation
of the projected angle on the sky plane and therefore a  combination
of the ejection angle, $\eta$, and  inclination.
The most powerful tool to determine the precession is the  radio astrometry:
following the shift of the peak of successive high resolution radio images 
we  determined  a period of $27-28$ d for the precession [27].
An independent estimate of the precession comes from the timing analysis.
In fact, precession implies a periodical Doppler boosting, 
that is periodical changes in the flux density
and this can be revealed  by  timing analysis.
A timing analysis of the 6.7 years of GBI radio data of \lsi has revealed indeed (Fig. 2) two rather close
frequencies: $P_1 = {1\over \nu_1} = 26.49 \pm 0.07$ days and
$P_2 = {1\over \nu_2} = 26.92 \pm 0.07$ days [28].
The period $P_1$ agrees with the value of
$26.4960 \pm 0.0028$ days [29] 
associated to the
orbital period of the binary system and corresponds to the predicted periodical accretion peak
along an eccentric orbit [30].
The period $P_2$ agrees well with the
estimate  by radio astrometry of $27-28$ days for the precession period.
Recently [31],
Lomb-Scargle analysis of \textit{Fermi}-LAT data around apoastron
revealed (Fig. 2)  the same  periodicities $P_{1_{\gamma}} = 26.48
  \pm 0.08$ d, $P_{2_{\gamma}} = 26.99 \pm 0.08$ d.
The similar behaviour of the emission at  high (GeV) and low (radio) energy towards apoastron 
is a hint for these emissions  to be caused by the same population of electrons in a precessing jet.

The beating of the two periodicities, $P_1$ and $P_2$,   likely gives rise to
the so called "long-term" modulation present in \lsp.
The peak flux density of the periodical radio outburst
exhibits  in fact a   modulation of 1667$\pm$8~d [29].
The beating of the two frequencies determined in the GBI timing analysis
gives straightforward the long-term modulation as  $1\over {\nu_1} - {\nu_2}$=1667 days [28].
Indeed,    a  physical
model [32] for \lsi{} of synchrotron emission from a precessing ($P_2$)
 jet, periodically ($P_1$) refilled with relativistic particles, produces
a maximum  (shown here in Fig. 4~a, b) when 
the jet electron density is at
  its maximum and the approaching jet forms the smallest
  possible angle with the line of sight. This coincidence of the
    highest number of emitting particles and the strongest Doppler
      boosting of their emission occurs with the frequency of
  $\nu_1-\nu_2$ creating the long-term modulation observed in
\lsp.
As one can see the model reproduces in fact also the minimum of the  log-term modulation
 (Fig. 4~c) corresponding at ejections when  the approaching jet forms the largest
 possible angle with the line of sight [32].

The  model  can determine the variation of the ejection angle, $\eta$,
at the epochs of   VLBI observations.
It is therefore of interest to  compare predicted variations of $\eta$ 
with the observed variations in position angle
of the radio structures in   high resolution images.
The trend of $\eta$
vs $\Phi (P_1)$ (orbital phase) is shown in Fig. 5  together with VLBA images [27,32].
 The plot of $\eta$ vs $\Phi$ reveals that
 run H was performed at the minimum $\eta$ and
D was performed  nearly at the  maximum $\eta$.
These two extreme situations of $\eta$ for the two runs, H and D, 
are sketched in Fig. 5 (right corner).
This important information   implies that
two runs at similar $\eta$, but one performed   before and
the other after run D,
refer to two jets pointing to opposite directions with respect to the axis
of the precession cone (Fig.~5, left corner).
If the model is correct, the position angle of the  associated radio structures
should reflect this different orientation.
Indeed, the  structure for run E points towards South-West, whereas
runs C and J show  structures pointing to South-East.
Similarly, the structure at run B points towards East whereas that for run F
points towards West.
Finally, for A, I, H, G our model results in a  low $\eta$ angle; this would correspond to a jet
 pointing closest to the earth (i.e., a micro-blazar). One can see that the related VLBA structures
develop indeed a North-South feature.
\begin{figure}[]
\center
\includegraphics[width=0.79\textwidth]{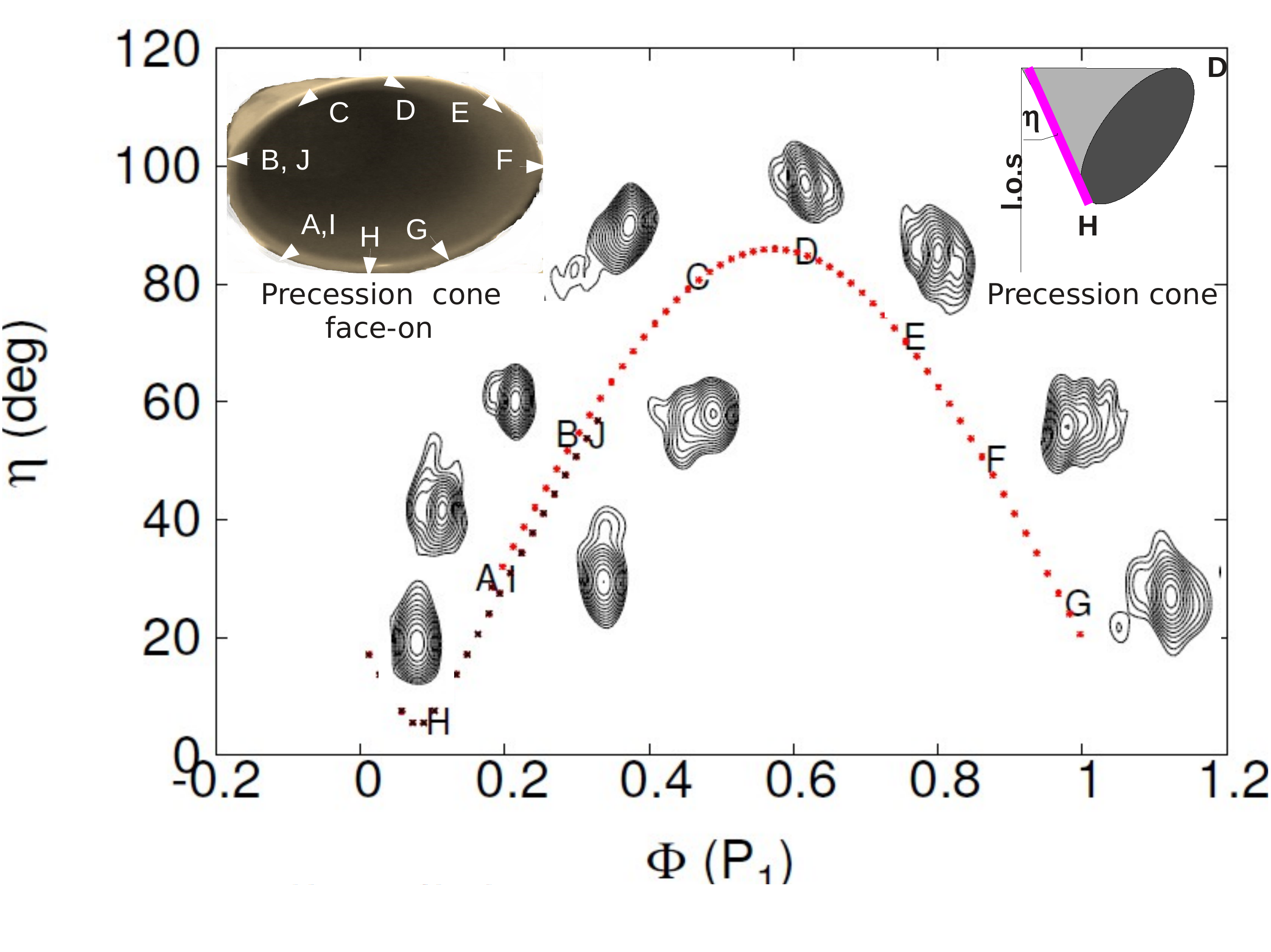}
\caption{
	VLBA images and related variations of the ejection angle $\eta$ because of precession (see Sect. 3.1) [32]
}
\end{figure}

%
To conclude this short review, 
there are different mechanisms producing transients in binary systems.
Intrabinary interaction 
of the magnetic structures of the two stellar objects in  \vt{}
is likely   the  origin of the transient in this system.
Variable accretion in an eccentric orbit
and consequent  refilling of a precessing jet is likely the cause for  the
modulated transient in the gamma-ray binary \lsp.
\acknowledgments
{\small
We thank Eduardo Ros and Frederic Jaron for  careful reading of the manuscript and helpful comments and suggestions.
This research is based on observations with the 100-m telescope of the
MPIfR (Max-Planck-Institut f\"ur Radioastronomie) at Effelsberg.
The Very Long Baseline Array is operated by the National Radio Astronomy Observatory, a facility of the National Science Foundation operated under cooperative agreement by Associated Universities, Inc.
This work has made use
of public \textit{Fermi}-LAT data obtained from the High Energy Astrophysics Science
Archive Research Center (HEASARC), provided by NASA Goddard Space
Flight Center.
}

%

\begin{thebibliography}{99}
\bibitem{}
  J. M. Cordes, T. J. W.  Lazio, M. A.  McLaughlin,
  \emph{The dynamic radio sky,
  New Astronomy Reviews} 
{\bf  48} (1459) 2004.
\bibitem{} A.~F. Boden, G. Torres, A.~I. Sargent, et al.,
\emph{Dynamical Masses for Pre-Main-Sequence Stars: A Preliminary
 Physical Orbit for V773 Tau A, ~ApJ} {\bf 670} (1214) 2007.
\bibitem{}
M. Massi, K. Menten, J. Neidh\"ofer,
\emph{Periodic radio flaring on the T Tauri star V 773 Tauri, ~A\&A},
{\bf 382} (152) 2002.
\bibitem{}
M. Massi, J. Forbrich, K. M. Menten, et al.,
\emph{Synchrotron emission from the T Tauri binary system V773 Tauri A,
A\&A} {\bf 453} 959 2006.
\bibitem{}
M. Nishio, K.  Yaji, T.  Kosugi, H. Nakajima, T.  Sakurai,
\emph{Magnetic Field Configuration in Impulsive Solar Flares Inferred from Coaligned Microwave/X-Ray Images, 
~ApJ}
{\bf 489} (976) 1997.
\bibitem{}
        Drake, M.A. Shay, W. Thongthai,  M. Swisdak,
\emph{Production of Energetic Electrons during Magnetic Reconnection,
  Physical Review Letters} {\bf 94} (95001) 2005.
\bibitem{}
M. Massi, E. Ros, K. M. Menten, et al.,
\emph{Interacting coronae of two T Tauri stars: first observational evidence for solar-like helmet streamers,
A\&A} {\bf 480} 489 2008.
\bibitem{}
M. Massi, E. Ros, D. Boboltz, et al.,
\emph{Interacting coronae of two T Tauri stars,
MemSAI} {\bf 84} 359 2013.
\bibitem{}
M. Massi, M. Kaufman Bernado', 
\emph{Magnetic field upper limits for jet formation,
A\&A} {\bf  477} 1 2008.
\bibitem{}
	E. B. Fomalont, B. J. Geldzahler, C. F. Bradshaw,
\emph{Scorpius X-1: The Evolution and Nature of the Twin Compact Radio Lobes 
Ap. J.} {\bf   558} 283 2001.
\bibitem{}
	V. Tudose, R.P. Fender, C.R. Kaiser, et al.,
\emph{The large-scale jet-powered radio nebula of Circinus X-1,
MNRAS} {\bf  372} 417 2006.
\bibitem{}
 R. E. Spencer, A. P. Rushton, M. Balucinska-Church, et al.,
\emph{Radio and X-ray observations of jet ejection in Cygnus X-2,
MNRAS} {\bf  435} L48 2013.
\bibitem{}
	R. P. Fender, T. M. Belloni,  E. Gallo.
\emph{Towards a unified model for black hole X-ray binary jets
MNRAS} {\bf  355} 1105 2004.
\bibitem{}
	D. Meier, S. Koide, Y. Uchida,
\emph{Magnetohydrodynamic Production of Relativistic Jets,
Science} {\bf  291} 84 2001.
\bibitem{}
C. Kaiser,
\emph{The flat synchrotron spectra of partially self-absorbed jets revisited,
MNRAS} {\bf  367} 1083 2006.
\bibitem{}
V. Dhawan, I. F. Mirabel, L. F. Rodriguez,
\emph{AU-scale synchrotron jets and superluminal ejecta in GRS 1915+105,
Ap. J.} {\bf  543} 373 2000.
\bibitem{}
	C. Brocksopp, S. Corbel, A. Tzioumis, et al.,
\emph{XTE J1752$-223$ in outburst: a persistent radio jet, dramatic flaring,
	multiple ejections and linear polarization,
MNRAS} {\bf  432} 931 2013.
\bibitem{}
T. W. Connors, S. Johnston, R. N. Manchester, D. McConnell,
\emph{The 2000 periastron passage of PSR B1259-63,
MNRAS} {\bf  336} 1201 2002.
\bibitem{}
	J. Mold\'on, S. Johnston, M. Rib\'o, J. M. Paredes, A. T. Deller,
\emph{Discovery of extended and variable radio structure from the gamma-ray binary 
	PSR B1259$-63$/LS 2883,
Ap. J.} {\bf  732} L10 2011.
\bibitem{}
J. Casares, I. Ribas, J.M. Paredes, J. Mart\'i, C. Allende Prieto
\emph{Orbital Parameters of the Microquasar LSI +61 303,
MNRAS} {\bf  360} 1105 2005.
\bibitem{}
M. Massi, M. Rib\'o, J. M. Paredes, et al.,
\emph{Hints for a fast precessing relativistic radio jet in LS I +61303,
A\&A} {\bf  414} L1 2004.
\bibitem{}
		V. Dhawan, A.  Mioduszewski, M.  Rupen,   
		\emph{LS I +61 303 is a Be-Pulsar binary, not a Microquasar},
\pos{PoS(MQW6)052} 2006.
\bibitem{}
P. C. Gregory, A. R.  Taylor, D. Crampton, et al.,
	\emph{The radio, optical, X-ray, gamma-ray star LSI +61 deg 303,
Astron. J.} {\bf 84} 1030 1979.
\bibitem{}
	L. Zimmermann, L. Fuhrmann, M. Massi,
\emph{The flat radio spectrum of \lsi and its evolution during outburst,
A\&A} {\bf  }  2015 submitted.
\bibitem{}
M. Massi, M. Kaufman Bernado', 
\emph{Radio Spectral Index Analysis and Classes of Ejection in LSI+61303,
A\&A} {\bf  702} 1 2009.
\bibitem{}
	D. F. Torres, N. Rea, P. Esposito, J. Li, Y. Chen, S. Zhang,
\emph{A Magnetar-like event from LS I +61303 and its nature as a gamma-ray binary,
Ap. J.} {\bf 744} 106 2012.
\bibitem{}
	M. Massi, E. Ros, L. Zimmermann,
\emph{VLBA images of the precessing jet of LS I +61303,
A\&A} {\bf  540} 142 2012.
\bibitem{}
M. Massi, F. Jaron,
\emph{Long-term periodicity in LS I +61303 as beat frequency between orbital and precessional rate,
A\&A} {\bf 554} 105 2013.
\bibitem{}
P. Gregory,
\emph{Bayesian Analysis of Radio Observations of the Be X-Ray Binary LS I +61303,
Ap. J.} {\bf 575} 427 2002.
\bibitem{}
V. Bosch-Ramon, J. M. Paredes, G. E. Romero, M. Rib\'o,
\emph{The radio to TeV orbital variability of the microquasar LS I +61303,
A\&A} {\bf 459} L25 2006.
\bibitem{}
F. Jaron, M. Massi
\emph{Discovery of a periodical apoastron GeV peak in LS I +61303,
A\&A} {\bf  572} 105 2014.
\bibitem{}
	M. Massi, G  Torricelli-Ciamponi, 
\emph{Intrinsic physical properties and Doppler boosting effects in LS I +61303,
A\&A} {\bf 564} 23 2014.
\end{thebibliography}
\end{document}